\documentclass{epl.mod}

\usepackage{psfrag}
\usepackage{epsfig}
\usepackage{amsmath,amssymb}
\usepackage{bm} 

\begin{document}

\title{Optical conductivity of a quasi-one-dimensional system with fluctuating order}
\shorttitle{Optical conductivity of a quasi-one-dimensional system \dots}
\author{Lorenz Bartosch}


\institute{Institut f\"{u}r Theoretische Physik, Universit\"{a}t Frankfurt, 
Robert-Mayer-Str. 8, Postfach 111932, 60054 Frankfurt, Germany}

\pacs{71.23.-k}{Electronic structure of disordered solids}
\pacs{71.45.Lr}{Charge-density-wave systems}
\pacs{72.80.-r}{Conductivity of specific materials}

\maketitle

\begin{abstract}
We describe a formally exact method to calculate the optical conductivity of a one-dimensional system with fluctuating order. For classical phase fluctuations we explicitly determine the optical conductivity by solving two coupled Fokker-Planck equations numerically. Our results differ considerably from perturbation theory and in contrast to Gaussian order parameter fluctuations show a strong dependence on the correlation length. 
\end{abstract}

%
%
%


In a strictly one-dimensional system with only short-range interactions order parameter fluctuations prevail over long-range order. There is no phase transition at any finite temperature. Although strictly one-dimensional systems do not exist in nature a large number of organic and inorganic compounds can be considered to be electronically quasi-one-dimensional. Their conductivity in a pronounced direction can be orders of magnitude larger than in any other direction.
An intensely studied class of such quasi-one-dimensional materials are Peierls systems.
Below a critical mean field temperature $T_c^{\rm MF}$ mean field theory predicts long-range order but one-dimensional fluctuations prohibit the occurrence of this symmetry broken state.
Only at temperatures much smaller than $T_c^{\rm MF}$ couplings between one-dimensional fluctuations become important and result in a finite critical transition temperature $T_c$ below which three-dimensional long-range order occurs as a charge density wave, a static lattice distortion, and a gap in the electronic spectrum \cite{Gruener94}. Precursors of the ordered phase can be seen far above the Peierls transition temperature $T_c$ (but below $T_c^{\rm{MF}}$) \cite{Johnston85} where one-dimensional fluctuating order dominates. A strong suppression of the density of states, {\em i.e.} a pseudogap, is quite similar to the pseudogap found in the cuprate superconductors and can also be observed in a number of correlation functions.

The instability of a Peierls system is due to the coupling of its conduction electrons to quasi-one-dimensional acoustic phonons. At not too small temperatures it is usually assumed that the phonons can be approximated by a static order parameter field $\tilde\Delta(x)$ which for an incommensurate lattice filling turns out to be complex. 
Above the phase transition the order parameter field can be considered to resemble disorder with short-range order such that its expectation value $\langle \tilde \Delta (x) \rangle$ has to be zero. The second moment decays exponentially, {\em i.e.}\ $\langle \tilde\Delta(x) \tilde\Delta^{\ast}(x^{\prime}) \rangle = \Delta_s^2 e^{-|x-x^{\prime}|/\xi}$. 
While the density of states of an incommensurate system is essentially unaffected by white noise ($\xi \to 0$, but $D \equiv \Delta_s^2 \xi=const$)\cite{Abrikosov78,Lifshits88} a pseudogap is established for sufficiently large correlation lengths $\xi$ \cite{Lee73,Sadovskii74,Sadovskii79,Lifshits88,McKenzie95a,McKenzie96b,Bartosch99d,Millis00b,Shannon00,Bartosch00d,Bartosch01b,Monien01,Bartosch03a}.
Numerical simulations show that for Gaussian statistics the density of states vanishes as $\rho(0) \propto \xi^{-\mu}$ (with $\mu \approx 0.64$) at the Fermi level \cite{Bartosch99d}. For classical phase fluctuations, however, it can be shown analytically that $\rho(0) \propto \exp(-4 \Delta_s \xi)$ \cite{Bartosch00d}. Hence, the finiteness of the correlation length and even the statistics of higher moments turn out to be very important. It is therefore obvious that a non-perturbative method which is not restricted to Gaussian colored noise should also be needed for an accurate calculation of one- or two-particle correlation functions.

Experimentally measurable transport properties can be related to two-particle correlation functions.
Of particular interest is the optical conductivity which is proportional to the optical absorption of a solid.
For a one-dimensional system with Gaussian white noise the optical conductivity
is given by the generally accepted Mott-Berezinskii law \cite{Mott79,Berezinskii73,Abrikosov78,Lifshits88,Fogler01}
\begin{equation}
  \label{eq:Mott}
  \sigma_1 (\omega) \sim 4 \sigma_0 (\omega \tau)^2 \ln^2(\omega \tau) \;.
\end{equation}
Here $\sigma_0$ is the Drude conductivity, $\omega$ is the frequency and $\tau$ is the elastic scattering time.
Of course it would be interesting to see how the Mott-Berezinskii law is modified by rendering the correlation length finite, opening also the possibility of studying non-Gaussian fluctuations.
The optical behavior of a system with fluctuating order would also be interesting within the light of optical investigations of the pseudogap phase of quasi-one-dimensional conductors like $($TaSe$_4)_2$I or K$_{0.3}$MoO$_3$ above their charge density wave transition \cite{Geserich86,Gruener94,Schwartz95,Perfetti02}.
The origin of the pseudogap observed in the optical conductivity of these systems can be considered to be complementary to the origin of a gap in a one-dimensional Mott insulator where the optical gap is dynamically generated by electron-electron interactions \cite{Jeckelmann00,Controzzi01}.

In this paper we will show how Halperin's method to calculate correlation functions for a particle in a white noise potential \cite{Halperin65,Lifshits88} can be generalized to calculate the optical conductivity of the fluctuating gap model which encapsulates the essential features of the pseudogap phase of quasi-one-dimensional Peierls systems and is also of interest for other systems \cite{Lee73,Lee74,Sadovskii74,Sadovskii79,Lifshits88,Bartosch01b}. The fluctuating gap model describes electrons with a linearized energy dispersion interacting with a static space-dependent order parameter field $\tilde \Delta(x)$ scattering electrons from one Fermi point to the other.
The non-perturbative calculation of the spectral function for this model with disorder determined by a generalized Ginzburg-Landau equation was addressed in a previous publication \cite{Bartosch03a}.

As the order parameter field $\tilde \Delta(x)$ is complex for an incommensurate lattice filling, fluctuations of $\tilde \Delta(x)$ can be subdivided into fluctuations of its amplitude $\Delta(x)$ and its phase $\vartheta(x)$, such that $\tilde \Delta(x) \equiv \Delta(x) e^{i\vartheta(x)}$. Using natural units which make Planck's constant $\hbar$, Boltzmann's constant $k_B$, and the Fermi velocity $v_F$ equal to unity the Hamiltonian of the fluctuating gap model is given by
\begin{equation}
  \label{eq:Hamiltonian}
  H = \left( 
  \begin{array}{cc}
    -i \partial_x & \tilde\Delta(x) \\
    \tilde\Delta^{\ast}(x) & i \partial_x 
  \end{array} \right) \;.
\end{equation} 
Since $H$ is invariant with respect to the charge conjugation operation $H \to \sigma_1 H^{\ast} \sigma_1$ where $\sigma_1$ is a Pauli matrix we can apply boundary conditions which imply that all eigenstates $\bm{\psi}_n (x)$ of $H$ are simultaneous even eigenstates of the charge conjugation operation $\bm{\psi}_n (x) \to \sigma_1 \bm{\psi}_n^{\ast} (x)$ such that $\bm{\psi}_n(x) = (\psi_n(x),\psi_n^{\ast}(x))^T$.


In the following we will be interested in the optical conductivity $\sigma_1(\omega)$ at finite temperature $T$
which can be expressed by the Kubo formula
\begin{equation}
  \label{eq:Kubosigma}
  \sigma_1 (\omega) = \pi \omega_p^2 \int_{-\infty}^{\infty} \upd\epsilon\, \frac{f(\epsilon) - f(\epsilon+\omega)}{\omega}\,F(\epsilon,\epsilon+\omega) \;.
\end{equation} 
Here $\omega_p^2 = 4\pi n e^2/m^{\ast}$ is the square of the plasma frequency (with $-e$ and $m^{\ast}$ the charge and effective mass of an electron and $n$ the density of electrons), $f(\epsilon) = [\exp(\epsilon/T)+1]^{-1}$ is the Fermi function and $F$ is defined as
\begin{equation}
  \label{eq:F}
  F(\epsilon,\epsilon^{\prime}) = \frac{1}{L} \left\langle \sum_{n,m} | J_{nm} |^2 \delta (\epsilon - \epsilon_n) \delta (\epsilon^{\prime} - \epsilon_m) \right\rangle \;, 
\end{equation}
where 
\begin{equation}
  \label{eq:J}
  J_{nm} = \int \upd x\, \bm{\psi}_n^{\dagger} (x) \sigma_3 \bm{\psi}_m (x)
\end{equation}
is a current matrix element.
The process of averaging in Eq.\ (\ref{eq:F}) restores translational symmetry. It is therefore possible to rewrite Eq.\ (\ref{eq:F}) as
\begin{equation}
  \label{eq:FII}
  F(\epsilon,\epsilon^{\prime}) = \bigg\langle \sum_{n,m} \int \upd x\, \psi_n^{\ast} (x) \psi_m(x) \psi_m^{\ast} (0) \psi_n (0) \, \delta (\epsilon - \epsilon_n) \, \delta (\epsilon^{\prime} - \epsilon_m) \bigg\rangle \;. 
\end{equation}
Let us now decompose the wave function $\psi_n(x)$ according to
\begin{equation}
  \label{eq:Eikonal}
  \psi_n(x) = i A e^{  \zeta_{\epsilon_n}(x)+i[\varphi_{\epsilon_n}(x) + \vartheta(x)]/2} \;.
\end{equation}
It should be noted that $\varphi_{\epsilon_n}(x)$ measures the phase of $\psi_n(x)$ relative to the phase of the order parameter field $\vartheta(x)$.
From $H \bm{\psi}_n(x) = \epsilon_n \bm{\psi}_n(x)$ it follows immediately that the phase variables $  \zeta_{\epsilon_n}$ and $\varphi_{\epsilon_n}$ satisfy the following equations of motion \cite{Bartosch01b},
\begin{eqnarray}
  \label{eq:phiII}
  \partial_x \varphi_{\epsilon}(x) & = & 2 \epsilon -  \partial_x \vartheta(x) + 2 \Delta(x) \cos\left[\varphi_{\epsilon}(x)\right] \;, \\
  \label{eq:zetaII}
  \partial_x \zeta_{\epsilon}(x) & = & \Delta(x) \sin\left[\varphi_{\epsilon}(x)\right] \;.
\end{eqnarray}
To construct all eigenstates for a given realization of the disorder potential we follow Ref.\ \cite{Lifshits88} and integrate the equations of motion (\ref{eq:phiII},\ref{eq:zetaII}) from the boundaries at $x = \mp L/2$ to $x=0$ where we try to match these solutions. If $\varphi_{\epsilon,\pm}(x)$ and $\zeta_{\epsilon,\pm}(x)$ are the solutions to Eqs.\ (\ref{eq:phiII},\ref{eq:zetaII}) when advancing the solution from $x=\mp L/2$ towards $x=0$, the matching condition which determines the discrete set of energies $\epsilon_n$ corresponding to the eigenstates $\psi_n(x)$ is 
\begin{equation}
  \label{eq:matching_condition}
  \varphi_{\epsilon,+}(0) = \varphi_{\epsilon,-}(0) \;.
\end{equation}
Instead of working with normalized wave functions which satisfy $\int \upd x\, \bm{\psi}_n^{\dagger} (x) \bm{\psi}_m (x) = \delta_{n,m}$ let us from now on use the normalization condition $|\psi_n(0)|=1$ which amounts to the replacement
\begin{equation}
  \label{eq:normalization}
  \psi_n(x) \to \frac{\psi_n(x)}{(2\int \upd x\, |\psi_n(x^{\prime})|^2)^{1/2}} \;. 
\end{equation}
Making use of the identity \cite{Lifshits88,Bartosch03a}
\begin{equation}
  \label{eq:deltaIII}
  \delta(\varphi_{\epsilon,+}(0) - \varphi_{\epsilon,-}(0)) = \sum_n \frac{\delta(\epsilon - \epsilon_n)}{2 \int_{-L/2}^{L/2} \upd x^{\prime}\, |\psi_n(x^{\prime})|^2} \;.
\end{equation}
which can be shown by using the theorem $\delta(f(x)) = \sum_n \frac{1}{|f^{\prime} (x)|} \delta(x-x_n)$ where the $x_n$'s are the zeros of $f$, we can now rewrite $F$ as
\begin{equation}
  \label{eq:FIII}
  F(\epsilon,\epsilon^{\prime})  =  2\, \rm{Re}\, \bigg\langle W_{+}(0)\, 
  \delta (\varphi_{\epsilon,+}(0)-\varphi_{\epsilon,-}(0)) \,
  \delta (\varphi_{\epsilon^{\prime},+}(0)-\varphi_{\epsilon^{\prime},-}(0))
 \bigg\rangle \;. 
\end{equation}
The quantities $W_{\pm}(x)$ are defined as
\begin{equation}
  \label{eq:W}
  W_{\pm}(x) \equiv \frac{1}{\psi_{\epsilon}^{\ast}(x) \psi_{\epsilon^{\prime}}(x)}
  \int_{\mp L/2}^{x} \upd x^{\prime} \psi_{\epsilon}^{\ast} (x^{\prime}) \psi_{\epsilon^{\prime}}(x^{\prime})
\end{equation}
and satisfy the equation of motion
\begin{equation}
  \label{eq:motion:W}
  \partial_x W(x) = -i\left[\epsilon^{\prime} - \epsilon + \Delta(x) e^{-i\varphi_{\epsilon^{\prime}}} - \Delta(x) e^{i\varphi_{\epsilon}}\right] W(x) + 1 \;.
\end{equation}

At this point let us specify the process of averaging. We assume that the probability distribution of $\{\tilde \Delta(x)\}$ can be expressed via a generalized Ginzburg-Landau theory, {\em i.e.}\ $P \{\tilde \Delta \} = \frac{1}{Z} e^{- F \{\tilde \Delta \}/T}$, where 
\begin{equation}
  \label{eq:Ginzburg_Landau}
  F \{\tilde \Delta\}/T = \int_{-L/2}^{L/2} \upd x \, \left[ \frac{M(\Delta(x))}{2} \left(\partial_x \Delta(x)\right)^2 + \frac{n(\Delta(x))}{2} \left(\partial_x \vartheta (x)\right)^2 + V(\Delta(x)) \right] \;.
\end{equation}
$M(\Delta)$, $n(\Delta)$, and $V(\Delta)$ can in principle be evaluated microscopically but can also be taken to be phenomenological given functions. At low temperatures, amplitude fluctuations get frozen out, {\em i.e.}\ $\Delta(x) \approx \Delta_s$ and Eq.\ (\ref{eq:Ginzburg_Landau}) reduces to a form describing classical phase fluctuations,
\begin{equation}
  \label{eq:phasefluctuations}
  F \{\vartheta\} = \frac{n_s}{2} \int_{-L/2}^{L/2} \upd x \,  \left(\partial_x \vartheta (x)\right)^2 \;.
\end{equation}
Here $n_s \equiv T \, n$ measures the stiffness of long-wavelength fluctuations and is approximately independent of temperature. A two-dimensional analogue to Eq.\ (\ref{eq:phasefluctuations}) has been used by Emery and Kivelson \cite{Emery95a} 
to explain the pseudogap phase in the underdoped cuprates.

The essential property of Eq.\ (\ref{eq:Ginzburg_Landau}) is that the stochastic process corresponding to Eq.\ (\ref{eq:Ginzburg_Landau}) has the Markov property and hence can be described by the Langevin equations
\begin{eqnarray}
  \label{eq:Delta}
  \partial_x \Delta (x) & = & a(\Delta)  + b(\Delta) \, \eta_{\Delta}(x) \;, \\
  \label{eq:theta}
  \partial_x \vartheta (x) & = & n^{-1/2} (\Delta) \, \eta_{\vartheta}(x) \;,
\end{eqnarray}
where $\eta_{i}(x)$, $i=\Delta,\vartheta$ are independent Gaussian white noise stochastic processes with $\langle \eta_{i}(x)\eta_{i}(x^{\prime})\rangle= \delta(x-x^{\prime})$. The functions
$a(\Delta)$ and $b(\Delta)$ may be obtained from Eq.\ (\ref{eq:Ginzburg_Landau}) using the transfer matrix method \cite{Scalapino72,Monien01,Bartosch03a}.
Decoupling the average in Eq.\ (\ref{eq:FIII}) by introducing the auxiliary variable $\Delta$ we obtain
\begin{equation}
  \label{eq:FIIII}
  F(\epsilon,\epsilon^{\prime}) = 2\, \rm{Re}\int_{0}^{\infty} \upd\Delta \int_{0}^{2\pi} \upd\varphi \int_{0}^{2\pi} \upd\varphi^{\prime}\,\frac{P_1(\varphi,\varphi^{\prime},\Delta;0) \,P_0(-\varphi,-\varphi^{\prime},\Delta;0)}{P(\Delta)} \;,
\end{equation}
where $P(\Delta) = \langle \delta(\Delta-\Delta(x)) \rangle$ and 
\begin{eqnarray}
  \label{eq:Pn}
    P_n(\Delta,\varphi,\varphi^{\prime};x) & = &
  \langle \left( W_{+} (x) \right)^n\, \delta(\Delta-\Delta(x)) \, \delta(\varphi-\varphi_{\epsilon,+}(x)) \, \delta(\varphi^{\prime}-\varphi_{\epsilon^{\prime},+}(x)) \rangle \;,\ n = 0,1\;
\end{eqnarray}
are joint probabilities.
Differentiating Eq.\ (\ref{eq:Pn}) with respect to $x$ we obtain the following coupled three-dimensional Fokker-Planck equations 
\begin{eqnarray}
  \label{eq:FP:P0}
    \partial_x P_0 & = &\Big[ \frac{(\partial_{\varphi} + \partial_{\varphi^{\prime}})^2}{2n} - 2 \partial_{\varphi} (\epsilon + \Delta\cos \varphi) - 2 \partial_{\varphi^{\prime}} (\epsilon^{\prime} + \Delta\cos \varphi^{\prime}) + \frac{(\partial_{\Delta} b)^2}{2}  - \partial_{\Delta} a \Big] P_0 \;,
\\
\label{eq:PDE:P1}
    \partial_x P_1 & = &\Big[ \frac{(\partial_{\varphi} + \partial_{\varphi^{\prime}})^2}{2n} - 2 (\epsilon + \Delta\cos \varphi) \partial_{\varphi} - 2 (\epsilon^{\prime} + \Delta\cos \varphi^{\prime}) \partial_{\varphi^{\prime}} 
\nonumber \\ 
& & \hspace{32mm}{} -i(\epsilon^{\prime} - \epsilon + \Delta e^{i\varphi^{\prime}} - \Delta e^{-i\varphi})  + \frac{(\partial_{\Delta} b)^2}{2}  - \partial_{\Delta} a \Big] P_1 + P_0 \;.
\end{eqnarray}
The solutions to these equations determine the exact optical conductivity for our problem and are similar to Halperin's equations to determine the optical conductivity of a particle in a white noise potential \cite{Halperin65,Lifshits88}. Unfortunately solving for the stationary solutions to Eqs.\ (\ref{eq:FP:P0},\ref{eq:PDE:P1}) is not an easy task. 
Both equations are parabolic partial differential equations. The absence of a diffusion term $(\partial_{\varphi} - \partial_{\varphi^{\prime}})^2$ makes their numerical solutions notoriously unstable. However, adding a small but finite diffusion term $\kappa (\partial_{\varphi} - \partial_{\varphi^{\prime}})^2$ turns the equations into elliptic partial differential equations which are numerically accessible. 
To further simplify our problem we will restrict ourselves to the physically especially interesting case of classical phase fluctuations where $2 n=\xi$ and $\Delta$ drops out of all equations. This reduces the dimensionality of the partial differential equations to two. Since the coefficients of both equations involve only exponentials of $\varphi$ and $\varphi^{\prime}$ and since both equations are subject to periodic boundary conditions the Fourier transform method seems to be the method of choice to solve for $P_0$ and $P_1$. 
We find that as we increase the number of Fourier components, the parameter $\kappa$ can be chosen smaller and smaller to obtain a stable solution. For sufficiently small $\kappa$ our solutions become independent of $\kappa$. 

\begin{figure}
  \begin{center}
    \psfrag{omega}{\large \hspace{5.5mm}$\omega/\Delta_s$}
    \psfrag{conductivity}{\large \hspace{3.8mm}$\sigma_1 (\omega)/(\Delta_s^{-1}\omega_p^2)$}
    \psfrag{0}{\large \hspace{0mm}$0$}
    \psfrag{0.1}[c][c]{\large $0.1$}
    \psfrag{0.2}[c][c]{\large $0.2$}
    \psfrag{1}[c][c]{\large $1$}
    \psfrag{2}[c][c]{\large $2$}
    \psfrag{3}[c][c]{\large $3$}
    \psfrag{4}[c][c]{\large $4$}
    \psfrag{0.5}[c][c]{\bf \small $0.5$}
    \psfrag{1.0}[c][c]{\bf \small $1$}
    \psfrag{2.0}[c][c]{\bf \small $2$}
    \psfrag{4.0}[c][c]{\bf \small $4$}
    \psfrag{8.0}[c][c]{\bf \small $8$}
    \psfrag{inf}[c][c]{\bf \small $\infty$}
    \epsfig{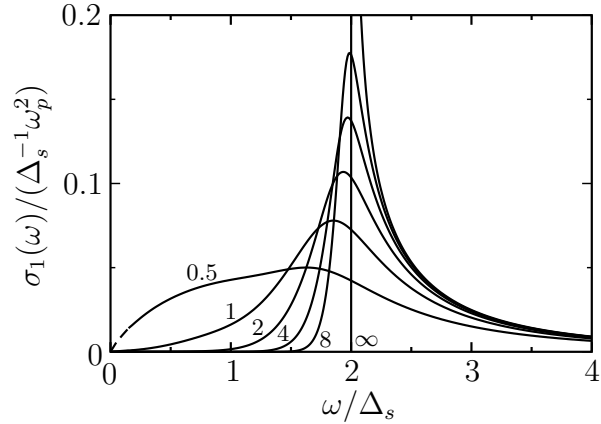} 
  \end{center}
  \caption{Optical conductivity calculated for classical phase fluctuations and $T=0$. The correlation length is given by $\Delta_s \xi = 0.5,1,2,4,8,$ and $\infty$. Dashed line: extrapolation between $\sigma_1 (0)=0$ and our numerical data for $\Delta_s \xi = 0.5$.}
  \label{fig:conductivity_Teq0}
\end{figure}

In Fig.\ \ref{fig:conductivity_Teq0} we show the optical conductivity calculated for various correlation lengths and zero temperature.
Unfortunately we have not been able to calculate $\sigma_1(\omega)$ for $\omega \lesssim 0.1 \Delta_s$.
The difficulty to calculate $\sigma_1 (\omega)$ for small $\omega$ can be understood as follows: As $\epsilon^{\prime}$ approaches $\epsilon$ the phase $\varphi_{\epsilon}(x)$ will approach $\varphi_{\epsilon^\prime}(x)$ such that $P_0(\varphi,\varphi^{\prime}) \to P_0(\varphi) \delta(\varphi - \varphi^{\prime})$. The singularity in this probability distribution obviously causes numerical difficulties.
As can be seen in Fig.\ \ref{fig:conductivity_Teq0} the restriction to frequencies $\omega \gtrsim 0.1$ does only imply that we cannot use our numerical methods to determine the quantitative corrections to the Mott-Berezinskii law. All curves aim towards $\sigma_1(0) = 0$ which is consistent with the fact that the dc conductivity of a one-dimensional disordered system should be zero.
In the limit $\xi \to \infty$ the optical conductivity approaches the well-known result of a Peierls insulator \cite{Lee74},
\begin{equation}
  \label{eq:conductivity:xi_to_infinity}
  \sigma_1 (\omega) = \frac{\omega_p^2}{2}\, \frac{\Delta_s^2 \theta(\omega^2 -4 \Delta_s^2)}{\omega^2 (\omega^2 - 4 \Delta_s^2)^{1/2}} \;,
\end{equation}
involving a square root singularity at $\omega \approx 2 \Delta_s$.
The asymptotic $1/\omega^3$-decay applies to all finite correlation lengths $\xi$ and shows that for large $\omega$ perturbation theory is accurate. 
In the pseudogap regime, however, $\sigma_1 (\omega)$ shows a strong dependence on the correlation length $\xi$. This is in contrast to the case of Gaussian order parameter fluctuations (for which there exists an ``almost exact'' solution that has also been applied to a two-dimensional model of the pseudogap state, see Ref.\ \cite{Sadovskii02}) where the dependence on $\xi$ is rather weak and no singularity arises for $\xi \to \infty$ \cite{Sadovskii74}.
Finally, as a nontrivial check for the optical conductivity we have the conductivity sum rule \cite{Pines66}
\begin{equation}
  \label{eq:sumrule}
  \int_0^{\infty} \upd\omega\, \sigma_1(\omega) = \frac{\omega_p^2}{8} \;,
\end{equation}
which is well satisfied by our numerics.

\begin{figure}
  \begin{center}
    \psfrag{omega}{\large \hspace{5.5mm}$\omega/\Delta_s$}
    \psfrag{conductivity}{\large \hspace{3.8mm}$\sigma_1 (\omega)/(\Delta_s^{-1}\omega_p^2)$}
    \psfrag{0}{\large \hspace{0mm}$0$}
    \psfrag{0.1}[c][c]{\large $0.1$}
    \psfrag{0.2}[c][c]{\large $0.2$}
    \psfrag{1}[c][c]{\large $1$}
    \psfrag{2}[c][c]{\large $2$}
    \psfrag{3}[c][c]{\large $3$}
    \psfrag{4}[c][c]{\large $4$}
    \psfrag{0.5}[c][c]{\bf \small $0.5$}
    \psfrag{1.0}[c][c]{\bf \small $1$}
    \psfrag{2.0}[c][c]{\bf \small $2$}
    \psfrag{4.0}[c][c]{\bf \small $4$}
    \psfrag{8.0}[c][c]{\bf \small $8$}
    \psfrag{inf}[c][c]{\bf \small $\infty$}
    \epsfig{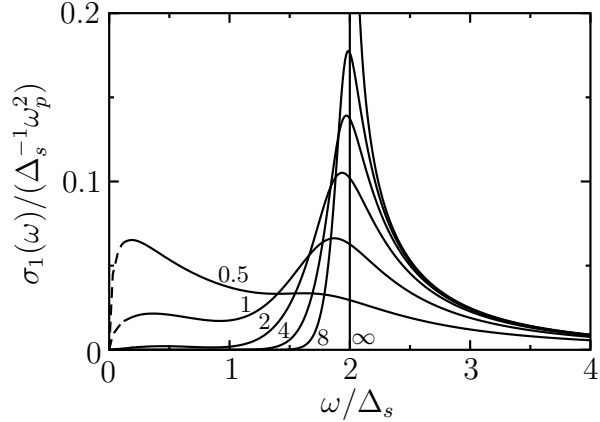} 
  \end{center}
  \caption{Optical conductivity calculated for classical phase fluctuations, $\Delta_s \xi = 0.5,1,2,4,8,\infty$ and temperature given by $T=1/\pi \xi$. Dashed lines: extrapolation between $\sigma_1 (0)=0$ and our numerical data.}
  \label{fig:conductivity_Tgr0}
\end{figure}

In the pseudogap phase of incommensurate Peierls systems the correlation length $\xi$ is approximately given by $\xi = 1/\pi T$ \cite{Gruener94,Bartosch00d}. We therefore show in Fig.\ \ref{fig:conductivity_Tgr0} the optical conductivity for different values of the correlation length and corresponding temperatures.
Due to thermal excitations, weight of the peak at $\omega \approx 2 \Delta_s$ is moved to lower frequencies. In this case the behavior of $\sigma_1(\omega)$ for small $\omega$ is not so obvious. 
However, we know that {\em all} eigenstates of our strictly-one-dimensional system are localized and therefore the dc conductivity should vanish for {\em any} finite temperature \cite{Lifshits88}. This enables us to find an extrapolation of $\sigma_1(\omega)$ in the frequency regime $0<\omega\lesssim 0.1$. We would like to note that our results differ from perturbative results of the optical conductivity which were calculated within the same model to first order in perturbation theory without considering vertex corrections \cite{Shannon00}. The perturbative results give a finite dc conductivity which for this model must be an artifact of perturbation theory but nevertheless explain the qualitative behavior of the dc conductivity of Peierls systems above their charge density wave transition.

In real Peierls systems dynamical effects become important below a crossover energy scale $\omega_{\ast} \ll \Delta_s$. 
If pinning is only weak, a sliding charge density wave can form and lead to an enhanced optical conductivity at low frequencies. This collective dynamic behavior cannot be described by our model.
Above the energy scale $\omega_{\ast}$, however, our model is applicable and describes the peak structure at $\omega \approx 2\Delta_s$.

In summary, we have presented an exact method based on the phase formalism to calculate the optical conductivity of a one-dimensional system with fluctuating order.
The optical conductivity can be expressed through the solutions to two coupled partial differential equations which for classical phase fluctuations we have solved numerically.
Our results correct previous perturbative results and give an explanation of the precursors of a single particle gap in the optical conductivity. To explain the experimentally observed collective features at low frequencies the dynamics of the order parameter field should be included. It would also be interesting to study corrections to the Mott-Berezinskii law for finite correlation lengths and different statistics of higher moments of the order parameter field. The governing equations for such a calculation are our Fokker-Planck equations.

\acknowledgements
I would like to thank Peter Kopietz and Hartmut Monien for helpful discussions.



\end{document}